\documentclass[12pt,epsf]{article}
\usepackage{amssymb,amsmath}
\usepackage{graphicx}
\usepackage{setspace}

\newcommand{\be}{\begin{equation}}
\newcommand{\ee}{\end{equation}}
\newcommand{\bea}{\begin{eqnarray}}
\newcommand{\eea}{\end{eqnarray}}
\newcommand{\bear}{\begin{eqnarray}}
\newcommand{\eear}{\end{eqnarray}}
\newcommand{\beas}{\begin{eqnarray*}}
\newcommand{\eeas}{\end{eqnarray*}}
\newcommand{\ba}{\begin{array}}
\newcommand{\ea}{\end{array}}
\newcommand{\nn}{\nonumber}

\newcommand{\tr}{\mathrm{Tr}}

\newcommand{\diag}{\mathrm{diag}}

\newcommand{\bee}{\be}
\newcommand{\eee}{\ee}

\newcommand{\nbox}{{\,\lower0.9pt\vbox{\hrule \hbox{\vrule height 0.2 cm \hskip 0.19 cm \vrule height 0.2 cm}\hrule}\,}}

\def\href#1#2{#2}

\begin{document}
\begin{titlepage}
\hfill
\vbox{
    \halign{#\hfil         \cr
           } 
      }  
\vspace*{20mm}
\begin{center}
{\Large \bf Rindler Quantum Gravity}

\vspace*{16mm}
Bart{\l}omiej Czech, Joanna L. Karczmarek, Fernando Nogueira, Mark Van Raamsdonk
\vspace*{1cm}

{
Department of Physics and Astronomy,
University of British Columbia\\
6224 Agricultural Road,
Vancouver, B.C., V6T 1W9, Canada}

\vspace*{1cm}
\end{center}
\begin{abstract}

In this note, we explain how asymptotically globally AdS spacetimes can be given an alternate dual description as entangled states of a pair of hyperbolic space CFTs, which are associated with complementary Rindler wedges of the AdS geometry. The reduced density matrix encoding the state of the degrees of freedom in one of these CFTs describes the physics in a single wedge, which we can think of as the region of spacetime accessible to an accelerated observer in AdS. For pure AdS, this density matrix is thermal, and we argue that the microstates in this thermal ensemble correspond to spacetimes that are almost indistinguishable from a Rindler wedge of pure AdS away from the horizon, but with the horizon replaced by some kind of singularity where the geometrical description breaks down. This alternate description of AdS, based on patches associated with particular observers, may give insight into the holographic description of cosmologies where no observer has access to the full spacetime.

\end{abstract}

\end{titlepage}

\vskip 1cm

\section{Introduction and summary}\label{introsection}

\begin{figure}
\centering
\includegraphics[width=0.25\textwidth]{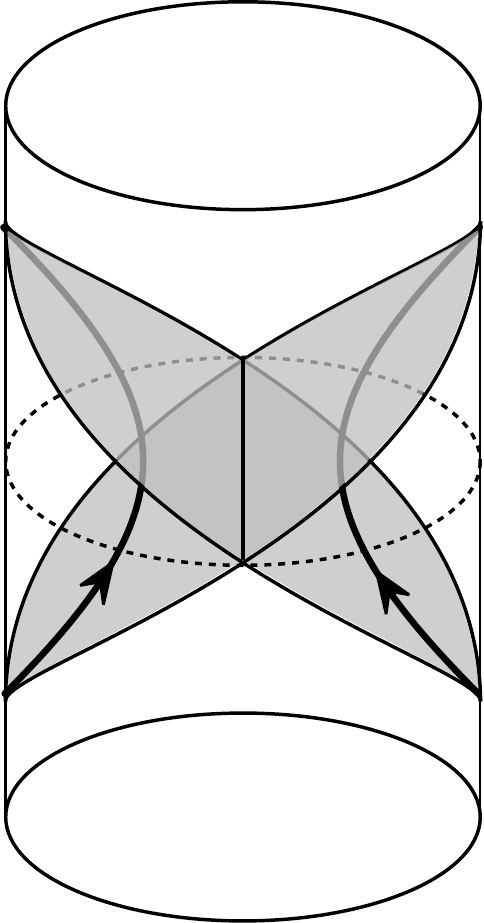}
\caption{A pair of accelerating observers in pure global AdS. The spacetime region accessible to each is a wedge whose boundary geometry can be chosen as $H^d \times R$. Each wedge has a dual description as a thermal state of a CFT on this $H^d \times R$ boundary geometry. The full spacetime is described by an entangled state of the two $H^d$ CFTs.}
\label{complementary}
\end{figure}

According to the AdS/CFT correspondence \cite{malda,agmoo}, asymptotically globally AdS spacetimes in certain quantum theories of gravity have an exact description as states of a conformal field theory on $S^{d}$. In this paper, we show (see Section~\ref{accobs}) that the same asymptotically AdS spacetimes may be described alternatively as entangled states of a pair of CFTs on hyperbolic space. This description in terms of hyperbolic space CFTs is precisely analogous to the description of Minkowski space field theory states in terms of entangled states of the field theory on two complementary Rindler wedges. In particular, if we focus on one of the $H^d$ CFTs, the degrees of freedom live in a density matrix, and this density matrix describes physics in a wedge of the dual spacetime accessible to an accelerated observer, as shown in Figure~\ref{complementary}.

The description of pure AdS in terms of the hyperbolic space theories is the specific entangled state\footnote{Here, and throughout this paper, we use $\sum$ to denote both discrete and continuous sums over states.}
\be
\label{fullstate}
|0_{global} \rangle = {1 \over Z} \sum_i e^{- \pi R_H E_i} |E^L_i \rangle \otimes |E^R_i \rangle \; ,
\ee
where $R_H$ is the curvature length scale of the hyperbolic space and $|E_i \rangle$ are energy eigenstates of the hyperbolic space CFTs. For this state, each hyperbolic space CFT is described by a thermal density matrix with temperature $(2 \pi R_H)^{-1}$, similar to the Rindler description of the Minkowski space vacuum.\footnote{The fact that the reduced density matrix associated with the boundary of a single wedge of pure AdS maps to a thermal density matrix for the CFT in hyperbolic space was demonstrated recently in \cite{Casini:2011kv}, which formed part of the inspiration for this work.}
State (\ref{fullstate}) has precisely the same form as the state of a pair of CFTs on $S^d$ that corresponds to the maximally extended eternal AdS-Schwarzschild black hole \cite{Israel:1976ur,Maldacena:2001kr}. This is no coincidence: thermal states of the $H^d$ CFT correspond to asymptotically AdS black holes with boundary geometry $H^d \times R$ \cite{emparan} and the choice of temperature $T = (2 \pi R_H)^{-1}$ is special in that it corresponds to a ``topological'' black hole that is locally pure AdS.

\begin{figure}
\centering
\includegraphics[width=0.6\textwidth]{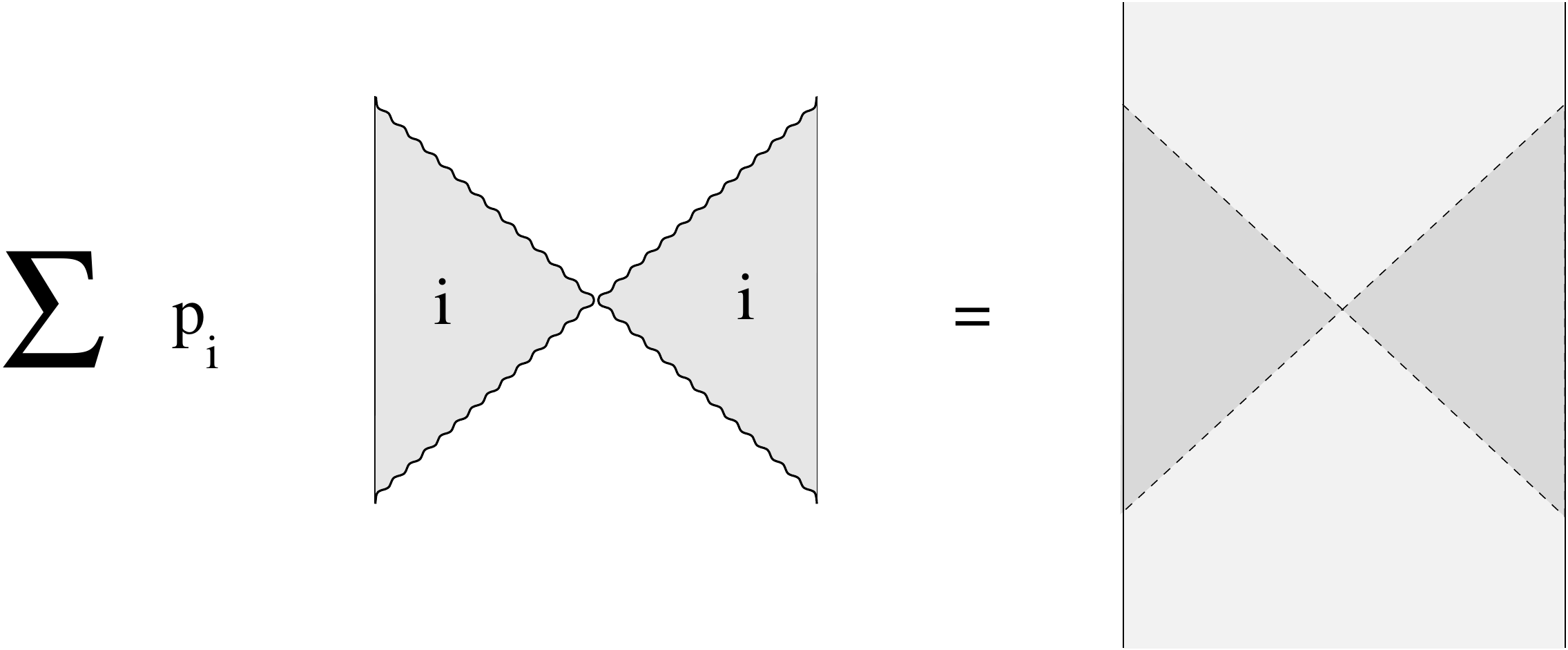}
\caption{Quantum superposition of microstate geometries yielding pure AdS spacetime. Each choice of complementary Rindler wedges leads to a different decomposition of AdS into a superposition of disconnected spacetimes.}
\label{superposition}
\end{figure}

If a Rindler wedge of pure AdS is described by a thermal density matrix, it is interesting to ask about the spacetime interpretation of the ``microstates'' contributing to this ensemble, i.e. the microstates of the topological black hole. We argue (see Sections~\ref{microstates} and \ref{rindler}) that typical pure states of the hyperbolic space CFT are dual to spacetimes that are almost indistinguishable from a Rindler wedge of pure AdS away from the horizon, but have the horizon replaced by some type of singularity where a geometrical description of the spacetime ceases to exist.\footnote{This similar to the ``fuzzball'' proposal of Mathur for black hole states; see \cite{Mathur:2008nj} for a review.} The description in equation~(\ref{fullstate}) of pure AdS may then be given a spacetime interpretation as in Figure~\ref{superposition}: a quantum superposition of disconnected singular wedges yields the connected global AdS spacetime.\footnote{This provides another explicit example of the idea \cite{VanRaamsdonk:2009ar,mvr2} that connected spacetimes emerge by entangling degrees of freedom in the non-perturbative description. Based on these observations, Mathur has argued \cite{Mathur:2010kx,Mathur:2011wg} that asymptotically flat spacetime could be understood as a quantum superposition of fuzzball geometries associated with Rindler microstates. Our AdS discussion here provides a concrete realization of Mathur's suggestions.} This description suggests strongly that the physics of AdS space outside the wedges (lightly shaded region in Figure~\ref{superposition}) is encoded in the information about how the two $H^d$ CFTs are entangled with each other.

The role of the microstate geometries is rather different for pure AdS as compared with an ordinary black hole. For black holes formed from collapse, the physical state is a pure state, one of the microstates of the black hole, and the black hole geometry may be understood as giving a coarse-grained description of the physics.  For pure AdS, the microstates have little to do with the physical spacetime. For these microstates, spacetime ends where the Rindler horizon would have been, while in the physical spacetime, the Rindler wedge is smoothly connected to a larger spacetime. The latter property is linked to the fact that the hyperbolic space CFT degrees of freedom are highly entangled with some other degrees of freedom. Thus, in describing pure AdS, it is crucial that the degrees of freedom of the hyperbolic space CFT are entangled with the other degrees of freedom, i.e. that they are genuinely described by a density matrix.

To highlight the importance of this entanglement, we consider in Section~\ref{disentanglement} a concrete realization of the ``disentangling experiment'' discussed in \cite{VanRaamsdonk:2009ar, mvr2}. Varying the temperature parameter in the state (\ref{fullstate}) away from $\beta = 2 \pi R_H$ changes the degree of entanglement between the two hyperbolic space CFTs (or the two halves of the sphere in the original picture) in a particular way. In this case, we can describe exactly what happens to the geometry: for any temperature $T$, the corresponding global geometry is the maximally extended hyperbolic space black hole with that temperature. From these explicit geometries, we can look specifically at what happens to a spatial slice of the spacetime as we vary the temperature. As the entanglement decreases, we find that the asymptotic regions corresponding to the two halves of the sphere become further apart and that the area of surfaces separating the two sides decreases, as argued on general grounds in \cite{VanRaamsdonk:2009ar, mvr2}.

\subsubsection*{Lessons for cosmological spacetimes}

\begin{figure}
\centering
\includegraphics[width=0.5\textwidth]{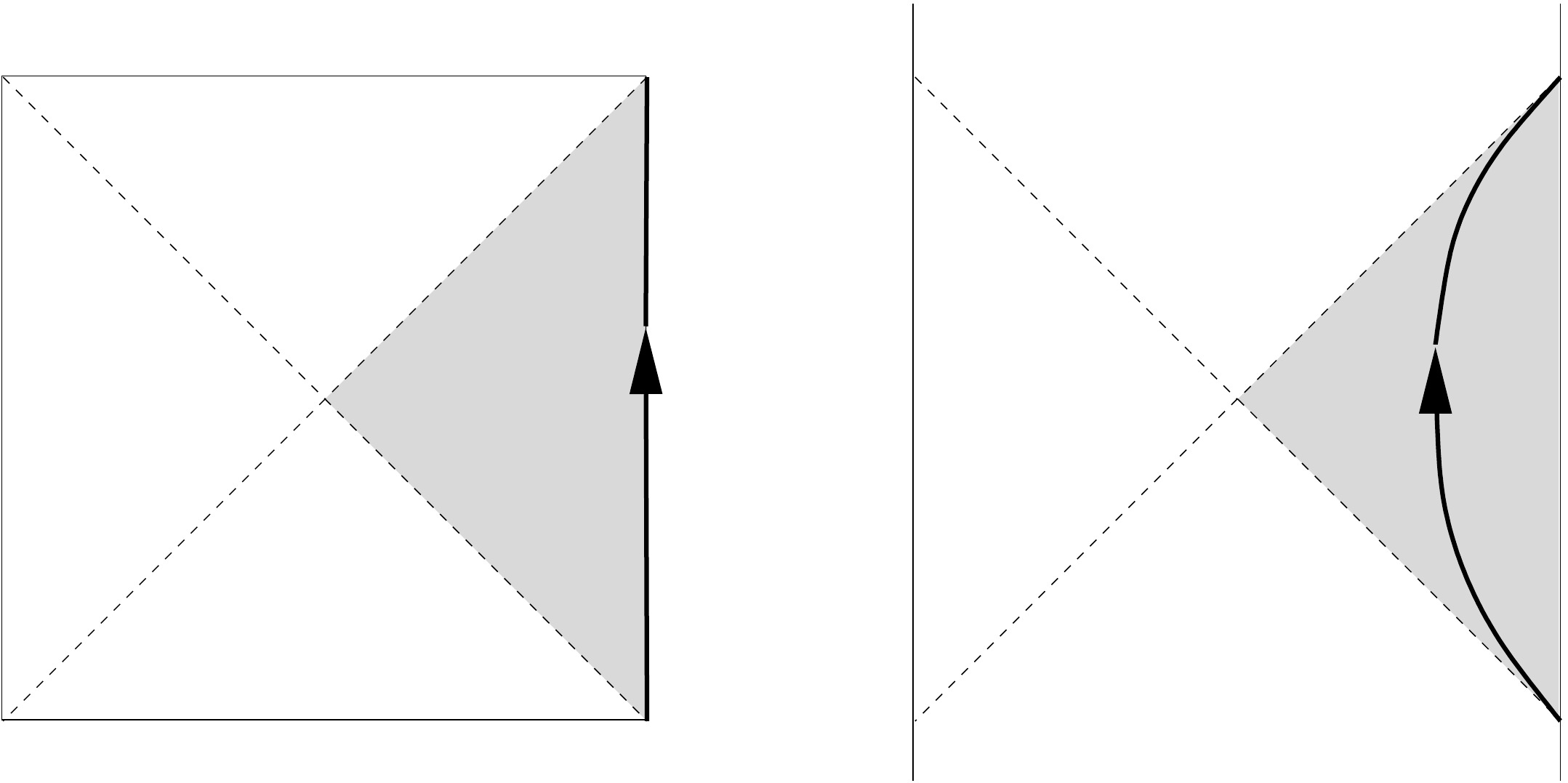}
\caption{Static observer in de Sitter space (left) and accelerated observer in AdS. Both have access to only a portion of the full spacetime, bounded (on one side in the AdS case) by a horizon with a thermal character.}
\label{dsads}
\end{figure}

The physics of accelerated observers in AdS spacetimes shares many qualitative features with the physics of observers in cosmological spacetimes with accelerated expansion. In Figure~\ref{dsads} (right), we have depicted an AdS observer with constant acceleration. The worldline for this observer starts and ends on the AdS boundary. This observer can communicate with (send light signals to and receive light signals back from) only a portion of the full global AdS spacetime shown by the shaded region in the figure. We see that this shaded region has a very similar character to the static patch accessible to a geodesic observer in de Sitter spacetime, shown on the left in Figure~\ref{dsads}. Both regions are bounded in the bulk spacetime by a horizon. Both observers see geodesic objects accelerating away from them towards the horizon. Finally, both horizons have a thermal character, emitting Unruh/Hawking radiation characteristic of some particular temperature.

These similarities give hope that some of the observations in this paper may be helpful in understanding how to generalize AdS/CFT to provide a non-perturbative description of quantum gravity in cosmological settings. In this context, it is interesting that we have given a precise description (via a density matrix for a subset of degrees of freedom) of patches accessible to particular observers in a complete model of quantum gravity. Like static patches in de Sitter space, these patches are bounded by observer-dependent horizons with an associated observer-dependent horizon area. In the de Sitter case, this observer-dependence obfuscates the proper interpretation of the entropy associated with this horizon area. In our present example, the interpretation of the observer-dependent entropy is clear: a single spacetime can be represented in many different ways as an entangled state of two subsets of degrees of freedom. Different choices of the subsets correspond to different patches (or different observers), and the the entropy associated to the horizon area in a patch measures the entanglement between the subsets. Alternatively, the entropy can be viewed as a count of microstates: the density matrix describing the subset of degrees of freedom associated with a single patch can be viewed as an ensemble of pure states, and each of these has a dual interpretation as a microstate geometry that is similar to the patch away from the horizon. It seems possible that all of these comments might apply equally well to de Sitter space or other cosmological spacetimes.\footnote{Our discussion here is similar to recent comments of Mathur in \cite{Mathur:2012zp}.} Some additional discussion on generalizing AdS/CFT to cosmological spacetimes is found in Section~\ref{cosmology}.

\section{A Rindler description of asymptotically global AdS spacetimes}
\label{accobs}

In the study of quantum fields on curved spacetime (or ``semiclassical'' quantum gravity), much of the physics of field theory on black hole backgrounds or on spacetimes with cosmological horizons can be understood by considering field theory on Rindler space, related to the physics experienced by accelerating observers in Minkowski space. It is interesting then to ask whether it is possible to describe precisely the physics accessible to an accelerated observer in a fully quantum mechanical description of gravity.
In this section, we shall see that for asymptotically globally AdS spacetimes described by states of a CFT on $S^d$, there is an alternate dual description that is precisely analogous to the Rindler description of field theory on Minkowski space.

\subsection{Asymptotically AdS spacetimes as entangled states of two hyperbolic space CFTs}

Consider an asymptotically globally AdS spacetime dual to a pure state $|\Psi_{S^d} \rangle$ of some CFT on $S^d \times R_t$. For any point $P$ on the boundary cylinder, we can consider the region $D_P$ consisting of all points on the boundary which are not timelike separated from $P$; for pure AdS, this forms the boundary of a Poincare patch. By a conformal transformation (reviewed in the appendix), the region $D_P$ can be mapped to Minkowski space; associated to this, we have a map from states of the $S^d$ CFT to states of the CFT on Minkowski space.\footnote{Since the region $D_P$ includes a complete spatial slice of the boundary cylinder (a boundary Cauchy surface) knowing the state of the fields on $D_P$ is the same information as knowing the fields on the entire boundary cylinder; thus, the map $|\Psi_{S^d} \rangle \to |\Psi_{R^d} \rangle$ is an isomorphism. Care must be taken in choosing the appropriate boundary conditions for the fields on Minkowski space.}

\begin{figure}
\centering
\includegraphics[width=0.9\textwidth]{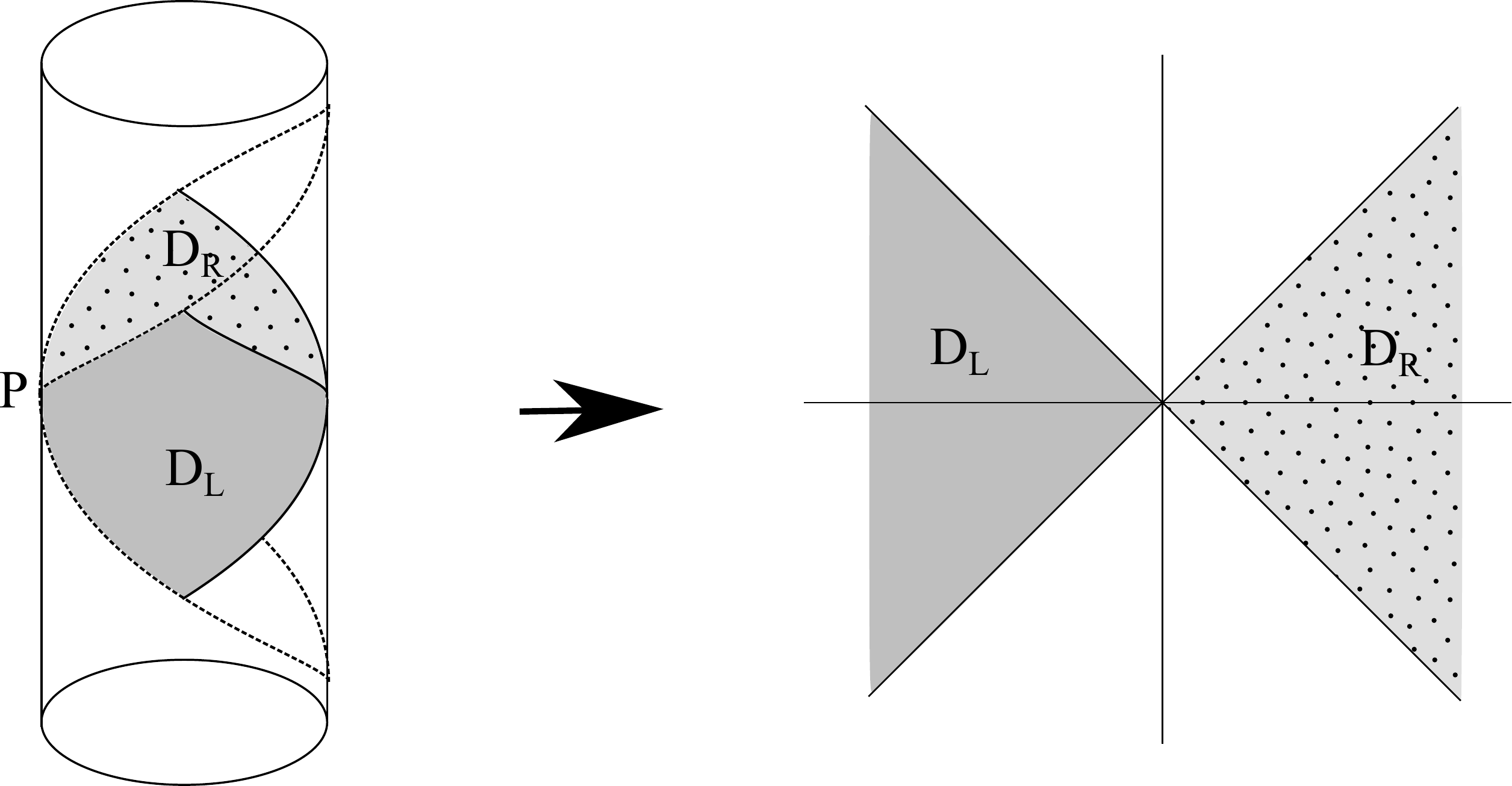}
\caption{Conformal map from the boundary of a Poincare patch to Minkowski space.
Region $D_L$ (solid) maps to one Rindler wedge of Minkowski space,
while the dotted region, $D_R$, maps to the other wedge.  The Poincare patch boundary $D_P$ is the region bounded by dashed lines.}
\label{Dtorind1}
\end{figure}

Now, consider two complementary Rindler wedges of Minkowski space, regions $R = \{x^1 \ge 0, |t| < x^1 \}$ and $L = \{x^1 \le 0, |t| < |x^1| \}$ for some choice of coordinates. Under the map from $D_P$, these regions are the images of two complementary ``diamond-shaped'' regions,\footnote{Each of these regions is the intersection of the interior of the future light cone of some point $p_i$ with the interior of the past light-cone of a point $p_f$ in the future of $p_i$. Alternatively, the regions are domain of dependence of a ball-shaped subset of some spatial slice of the boundary cylinder.} as shown in Figure~\ref{Dtorind1}. Any state of the CFT on Minkowski space can be represented as an entangled state of the quantum field theories on the separate Rindler wedges $R$ and $L$. For example, the Minkowski space vacuum is described in terms of field theories on the complementary Rindler wedges by the entangled state
\be
\label{full1}
|0_M \rangle = {1 \over Z} \sum_i e^{- \frac{\beta E_i}{2}} |E^L_i \rangle \otimes |E^R_i \rangle \; .
\ee
where $|E^L_i \rangle$ and $|E^R_i \rangle$ are corresponding eigenstates of the Rindler Hamiltonians on the two wedges (boost generators in the full Minkowski space).

By another conformal transformation (reviewed in the appendix) the Rindler wedges $R$ and $L$ can each be mapped to $H^d \times {\rm time}$, where $H^d$ is the $d$-dimensional hyperbolic space with metric
\begin{equation}
ds^2 = du^2 + R_H^2 \sinh^2\!\! \frac{u}{R_H} \,d\Omega_{d-1}^2 \; .
\end{equation}
Thus, the entangled state of the field theory on two Rindler wedges maps to an entangled state of the pair of CFTs on hyperbolic space. Under the conformal transformations to $H^d \times R_t$, the Rindler Hamiltonian in each wedge maps to the Hamiltonian generating time translations in $H^d \times R_t$. Thus, the state (\ref{full1}) describing pure global AdS spacetime maps to the state
\be
\label{full}
|0 \rangle_{S^d} = {1 \over Z} \sum_i e^{- \pi R_H E_i} |E^L_i \rangle_{H^d} \otimes |E^R_i \rangle_{H^d} \; .
\ee
of the pair of CFTs on $H^d \times R$. Here, one finds that the temperature parameter $\beta$ takes on the specific value $2 \pi R_H$.

In the CFT on $S^d$, the state corresponding to pure global AdS spacetime is clearly quite special: it is the energy eigenstate of the CFT Hamiltonian with the lowest possible energy. In the alternate description, the state (\ref{full}) is not in any sense a minimum energy state for the Hamiltonian of either $H^d$ CFT. However, states of the form (\ref{full}) have the maximum amount of entanglement entropy for a given energy expectation value.\footnote{This is true for the state (\ref{full}) at any temperature, but within this set of states, the one with $T = (2 \pi R_H)^{-1}$ is the only one with asymptotically global AdS asymptotics, as we will see explicitly in Section~\ref{rindler}.}

We will see below that states of the $H^d \times H^d$ CFT without entanglement correspond to states of the $S^d$ CFT with a singular stress-energy at the lightlike boundaries of the two regions that map to the two hyperbolic spaces. By the AdS/CFT dictionary, singularities in the stress-energy tensor can be associated with deformations of the metric which violate the asymptotically AdS boundary conditions. Thus, while we can associate a state of the $H^d \times H^d$ CFT to any asymptotically globally AdS spacetime (for a theory dual to a CFT on $S^d$), general states of the $H^d \times H^d$ CFT correspond to a more general class of spacetimes.

\subsection{The description of a single Rindler wedge}

So far, we have shown that any state of a CFT on $S^d \times R_t$ can be represented as an entangled state of a pair of hyperbolic space CFTs. This gives an alternate description of asymptotically global AdS spacetimes. We will now see that the information contained in the individual hyperbolic space CFTs corresponds (roughly) to the information accessible to a pair of complementary accelerating observers in the bulk. Thus, we can think of the hyperbolic space picture as giving a Rindler description of asymptotically global AdS spacetimes.

\begin{figure}
\centering
\includegraphics[width=0.25\textwidth]{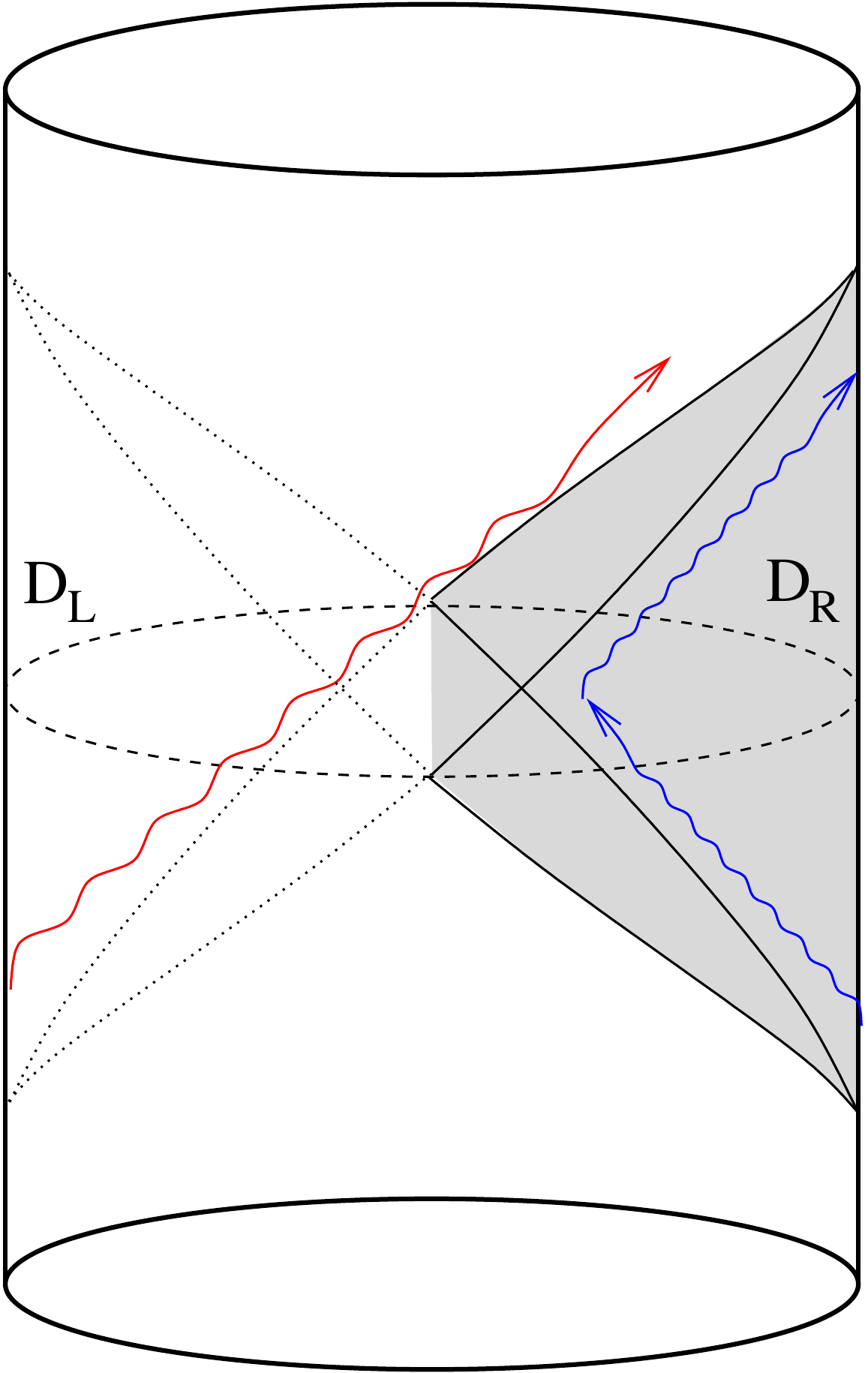}
\caption{Wedges of pure AdS spacetime. Field theory observables in $D_R$ (the shaded part of the boundary cylinder) probe the bulk region $J^+(D_R) \cap J^-(D_R)$ (the shaded region of the bulk). Any point in this region can receive a light signal (blue line) from and send a light signal to $D_R$. Physics outside this region can be altered by changes on the boundary that do not affect the state of the fields in $D_R$. One trajectory along which such changes propagate is shown in red.}
\label{density}
\end{figure}

Consider first the case of pure global AdS, described in the hyperbolic space CFT picture as the entangled state (\ref{full}). In this state, the reduced density matrix for each individual CFT is the thermal density matrix corresponding to temperature $1/(2 \pi R_H)$.\footnote{This has been derived directly in \cite{Casini:2011kv}.} Generally, thermal states of a CFT on hyperbolic space are dual to asymptotically locally AdS black hole solutions with boundary geometry $H^d$, discussed in detail in \cite{emparan} and reviewed in Section~\ref{disentanglement} below.\footnote{There is no analog of the Hawking-Page transition here, though we have a qualitative change in the causal structure of the maximally extended solutions at $T=1/(2 \pi R_H)$. For temperatures below $(2 \pi R_H)^{-1}$ the black holes have a causal structure similar to Reissner-Nordstrom AdS black holes while for temperatures higher than $(2 \pi R_H)^{-1}$ the causal structure is similar to Schwarzschild AdS black holes.} However, for the particular temperature $1/(2 \pi R_H)$, such a black hole is locally pure AdS. For such a black hole, the region outside the horizon corresponds exactly with a ``Rindler wedge'' of pure AdS, the region accessible to an accelerating observer whose worldline starts and ends at the past and future tips of the diamond-shaped region that maps to hyperbolic space, as shown in Figure~\ref{density}.\footnote{Here, an ``accessible'' point is one for which the observer can send a light signal to and receive a light signal back from that point.}

We will now argue that the density matrix for the single $H^d$ CFT generally does not carry any information about the region beyond this wedge. We recall that the two copies of $H^d \times R_t$ are related by conformal transformations to two complementary diamonds (which we refer to as $D_R$ and $D_L$) on the cylindrical boundary of global AdS.  For the full CFT on $S^d$, there are many pure states that give rise to precisely the same density matrix for the region $D_R$. For these states, the fields in the complementary region $D_L$ generally differ, and such differences can affect any point in the bulk in the causal past or causal future of $D_L$, as we see in Figure~\ref{density}. Thus, there are many states of the full field theory for which the density matrix for $D_R$ is the same but the physics in the region $J^+(D_L) \cup J^-(D_L)$ differs. We conclude that the density matrix associated with the region $D_R$ (equivalently, the density matrix for the corresponding $H^d$ CFT) knows only about the complement of $J^+(D_L) \cup J^-(D_L)$. But for pure AdS, this is exactly the region $J^+(D_R) \cap J^-(D_R)$ outside the horizon of the hyperbolic black hole with temperature $1/(2 \pi R_H)$.

To summarize, the density matrices for the two hyperbolic space CFTs describe the physics in two complementary Rindler wedges of pure AdS. The individual density matrices generally do not have information about the spacetime regions beyond the respective Rindler horizons. The additional information that comes from knowing the full state as compared with knowing the two density matrices is the information about how the degrees of freedom in the two CFTs are entangled with each other. Thus, we can say that the physics in the region outside the two Rindler wedges is described by the entanglement between the two hyperbolic space CFTs.\footnote{The arguments in this section apply equally well to Schwarzschild-AdS spacetimes dual to an entangled state of two $S^d$ CFTs. They suggest that a single $S^d$ CFT in a thermal density matrix knows only about the region outside the horizon of the black hole. Knowing anything about physics behind the horizon requires knowledge of both CFTs.}

In this section, we have focused on the case of pure global AdS spacetime. More general asymptotically AdS spacetimes correspond to other entangled states of the two hyperbolic space CFTs. The question of what region of these spacetimes is associated with the density matrix for a single hyperbolic space CFT (and more generally, what region of the spacetime dual to a state $|\Psi \rangle$ of a CFT on $M$ can be reconstructed from the density matrix for any spatial subset of degrees of freedom of the CFT) was considered by the present authors recently in \cite{Czech:2012bh} (and by others in \cite{Bousso:2012sj,Hubeny:2012wa}). There, we argued that the identification of the wedges $J^+(D_R) \cap J^-(D_R)$ and $J^+(D_L) \cap J^-(D_L)$ as the duals of the density matrices associated with $D_R$ and $D_L$ is somewhat specific to pure AdS. For more general spacetimes (with matter), the causal wedges $J^+(D_R) \cap J^-(D_R)$ and $J^+(D_L) \cap J^-(D_L)$ do not intersect in the bulk, but the density matrices associated with $D_R$ and $D_L$ carry information about larger wedges $w(D_R)$ and $w(D_L)$ that generally do intersect.

\section{The microstates of a Rindler wedge of AdS}
\label{microstates}

The state (\ref{full}) that describes pure AdS in the $H^d \times H^d$ picture has exactly the same form as the states of a CFT on $S^d \times S^d$ that describe maximally extended Schwarzschild-AdS black hole spacetimes \cite{Israel:1976ur,Maldacena:2001kr}. Specifically, in both cases the degrees of freedom of the two CFTs are entangled such that each ``factor'' theory is in a thermal state. The basic reason for this similarity is that (as we have seen) pure global AdS itself can be understood as a particular type of black hole.

We have argued that the thermal density matrix for a single CFT is dual to the region outside the horizon of the black hole. As usual, the (suitably regularized) area of the black hole horizon can be identified with the (regularized) entropy of the density matrix. For the full theory with a second CFT on $H^d$, this entropy would naturally be viewed as an entanglement entropy, measuring the extent to which the degrees of freedom are entangled with each other. But in discussions of black hole physics, such an entropy is more commonly viewed as a thermodynamic entropy counting microstates contributing to the ensemble described by the density matrix. Since the ``black hole'' in our case is a patch of pure AdS spacetime, it may seem odd to talk about its microstates. However, the thermal density matrix for the $H^d$ CFT can certainly be viewed as an ensemble of pure states and AdS/CFT suggests that these pure states should have some dual gravity interpretation. The goal of this section is to understand better these microstate geometries.

\subsubsection*{Interpreting the hyperbolic black hole microstates}

Recall that the $H^d$ CFT can be viewed as the theory on a Rindler wedge of a Minkowski space forming the boundary of a Poincare patch. Pure AdS corresponds to the Minkowski space vacuum and, as usual, the description of the fields on one Rindler wedge is via a thermal state. In this picture, the microstates are pure energy eigenstates of the Rindler space field theory, most of which are typical states in the ensemble described by the density matrix. For such typical states, we expect that almost any macroscopic observable will be nearly identical to the corresponding observable in the thermal state. Field theory observables (e.g. correlation functions, Wilson loops, entanglement entropies) tell us about the geometry of the dual spacetime, which suggests that the gravity dual of one of these typical microstates should be almost identical to the gravity dual of the thermal state itself. Let us try to understand this in more detail.

\begin{figure}
\centering
\includegraphics[width=1\textwidth]{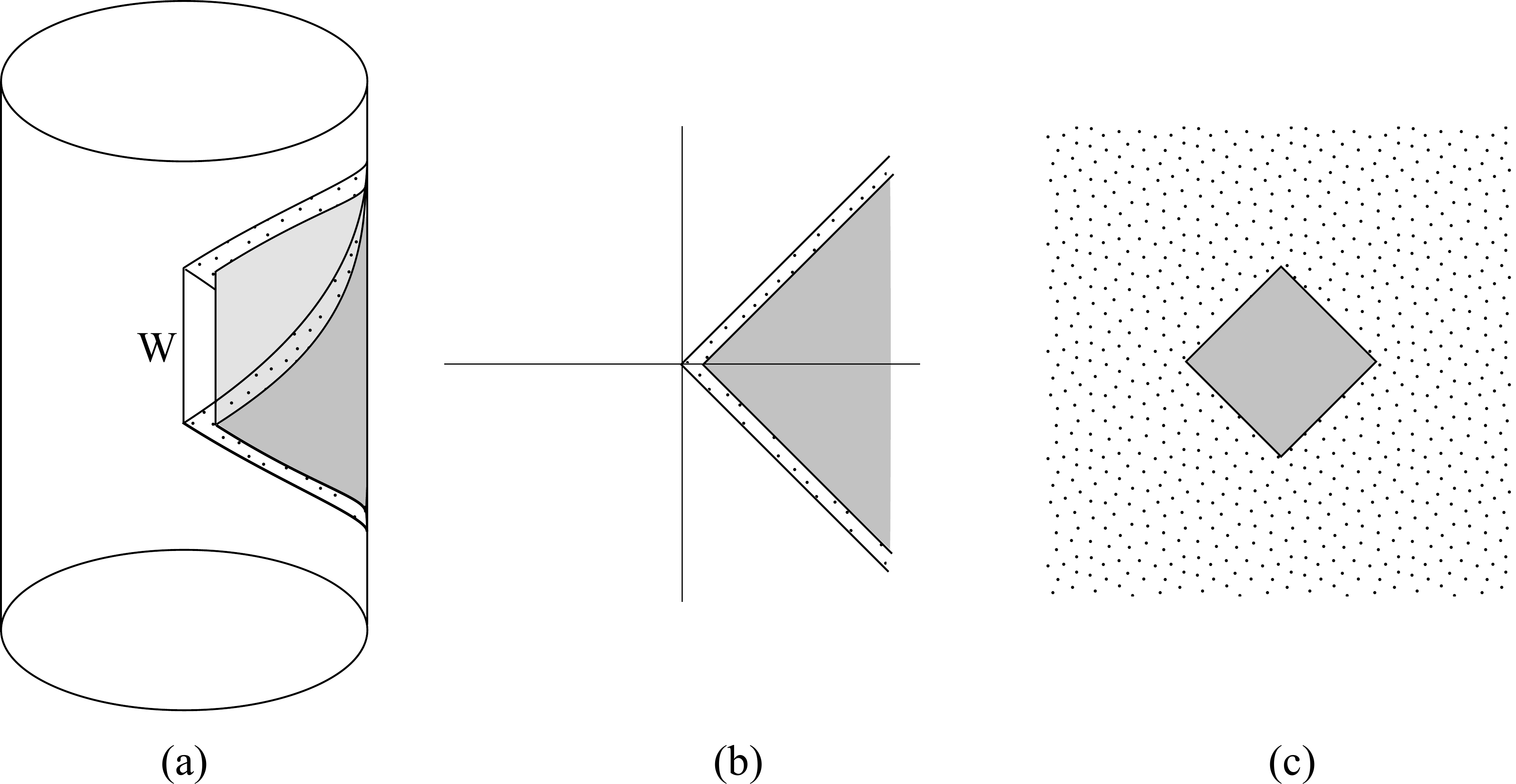}
\caption{Subregion $D_\epsilon$ (shaded) of boundary $D$ of a Rindler wedge of AdS.  The complement
of $D_\epsilon$ in $D$  is shown dotted.  (a) Regions $D$ and  $D_\epsilon$ on the boundary of AdS.
The corresponding bulk regions are also shown.  $W$ is a surface in the bulk whose
area computes the entanglement entropy of the fields in $D$.  (b) $D_\epsilon$  is mapped to
portion of a Rindler wedge.  (c)  $D_\epsilon$ is mapped to a finite portion of the
infinite hyperbolic plane.}
\label{interior}
\end{figure}

\subsubsection*{Microstate geometries look like the Rindler wedge of pure AdS away from the horizon}

Consider a domain of dependence region $D_\epsilon$ in the boundary spacetime,\footnote{The ``domain of dependence'' of a spatial region $A$ is the set of all points $p$ such that every inextensible causal curve through $p$ passes through $A$. The boundary $D$ of a Rindler wedge of pure AdS is the domain of dependence of a ball-shaped subset of a spherical constant-time slice of the boundary cylinder (e.g. a hemisphere of the $t=0$ sphere). The region $D_\epsilon$ can be taken as the domain of dependence of a slightly smaller ball.} which is slightly smaller than the boundary $D$ of our Rindler wedge (see Figure~\ref{interior}).
In the map from $D$ to hyperbolic space times time, the region $D_\epsilon$ maps to a finite region of $H^d \times R$. Thus, we can think of the degrees of freedom in $D_\epsilon$ as forming a small subset of the full set of degrees of freedom in the hyperbolic space field theory. For a typical microstate in some thermodynamic ensemble, the reduced density matrix for a small subset of degrees of freedom should be nearly the same as the reduced density matrix that arises from the thermal state itself \cite{canonical1,canonical2}. In fact, given the exact reduced density matrix $\rho_{D_\epsilon} (T)$ arising from the thermal state of the hyperbolic space CFT, there should be many pure states of the full theory for which the reduced density matrix on $D_\epsilon$ is exactly $\rho_{D_\epsilon} (T)$. The reason is that for any density matrix $\rho = \sum p_i |A_i \rangle \langle A_i |$ we can always choose a pure state $|\Psi \rangle = \sum p_i |A_i \rangle \otimes | B_i \rangle$ in a theory with a sufficiently large number of added degrees of freedom such that the reduced density matrix for the smaller system is exactly $\rho$. Here, we certainly have enough degrees of freedom, since there is an infinite volume of hyperbolic space outside the finite region that is the image of $D_\epsilon$ under the conformal transformation from $D$ to $H^d \times R$.\footnote{Note, however, that if a UV cutoff is imposed on the original field theory on $S^d$, there will be a limit to how large the region $D_\epsilon$ can be such that we can still choose a pure state on $D$ to exactly reproduce the density matrix $\rho_{D_\epsilon}$ arising from the vacuum of the $S^d$ CFT.} Thus, restricting to any subregion $D_\epsilon$ of $D$, there is no way in general that we can distinguish a pure microstate on $D$ from the mixed state on $D$ dual to pure AdS. According to \cite{Bousso:2012sj,Czech:2012bh,Hubeny:2012wa}, this means that the bulk region associated with the boundary $D_\epsilon$ will be the same as for pure AdS.\footnote{For a general microstate, the density matrix $\rho_{D_\epsilon}$ will not necessarily be exactly the same as the one arising from the thermal state, but typically it will be almost identical. The corresponding bulk region should then be almost indistinguishable from pure AdS.}

\subsubsection*{Rindler horizon replaced by something singular  in microstate geometries}

Let us now consider what happens at the boundary of the region $D$. After a conformal transformation that takes the region $D$ to a Rindler wedge of Minkowski space, this boundary becomes the Rindler horizon. For any pure state of quantum field theory on Rindler space (considered together with a pure state of the quantum field theory on the complementary Rindler wedge such that we have some state of the full Minkowski space field theory), we expect that the stress-energy tensor is singular on the Rindler horizon. This was shown for the state $|0_L \rangle \otimes |0_R \rangle$ of a free scalar field theory in \cite{parentani} and we demonstrate it more generally for any product state $|\Psi_L \rangle \otimes |\Psi_R \rangle$ in Section~\ref{rindler}. We also show in Section~\ref{sdhd} that the state $|0_L \rangle \otimes |0_R \rangle$ gives a singular stress-energy tensor on the Rindler horizon for any conformal field theory. We expect this conclusion to extend to any product state.\footnote{This is consistent with a result from algebraic quantum field theory that finite subregions of a quantum field theory do not admit pure states \cite{AQFT1,AQFT2}. This implies that starting from a product state in e.g. a lattice regularized theory, and taking a continuum limit must lead to some singular behavior at the interface between the regions.} Thus, while the field theory observables for a pure state on $D$ can be arbitrarily close to the vacuum observables away from the Rindler horizon, the behavior at this horizon (i.e. the boundary of $D$) is drastically different.

What is the bulk interpretation of this? We have seen above that the bulk region of the microstate spacetime associated with any smaller region $D_\epsilon$ will be almost indistinguishable from a wedge of pure AdS. On the other hand, as $D_\epsilon$ grows to become the full region $D$, drastic differences appear, with various observables diverging as we hit the boundary of $D$. This suggests that the bulk spacetime dual to a typical microstate of the $H^d$ CFT in the  $T = (2 \pi R_H)^{-1}$ thermal ensemble differs significantly from the Rindler wedge of pure AdS at the horizon. These differences can only occur at the horizon of the Rindler wedge region, since we have seen that any smaller wedge should be almost indistinguishable from AdS. This suggests that the dual of a microstate of the $T = (2 \pi R_H)^{-1}$ hyperbolic black hole should look like a Rindler wedge of AdS, but with the bulk horizon replaced by some type of singularity, probably of a non-geometric character.

More evidence for such singular behavior comes from considering the behavior of entanglement observables in the field theory and their proposed gravity dual description. We recall that according to the proposal of Ryu and Takayanagi \cite{Ryu:2006bv} (and the covariant generalization \cite{Hubeny:2007xt}), the von Neumann entropy of the density matrix associated with some spatial region $A$ of a quantum field theory with weakly curved holographic dual is equal to the area of an extremal surface $W$ in the dual spacetime such that the boundary of $W$ is the same as the boundary of $A$ (as in figure \ref{interior}). For a microstate of the $H^d$ CFT, we expect that the von Neumann entropy associated with a smaller region $D_\epsilon$ should be very similar to that for the same region in the thermal state. On the other hand, for the entire region $D$, the von Neumann entropy for the thermal state dual to a wedge of pure AdS is equal to the area of the Rindler horizon, while the von Neumann entropy of a pure microstate is zero. For the microstate, this implies either that the surface $W$ has zero area, or that the Ryu-Takayanagi formula no longer applies (e.g. because the relevant region of spacetime no longer has a weakly curved geometrical description). In either case, it appears that the metric on the boundary of the Rindler wedge is replaced by something singular (or non-geometric) when we pass from the black hole geometry (i.e. the wedge of pure AdS) to the microstate geometry.

\subsubsection*{Connections to previous work}

Our observations here illustrate and elaborate on an observation in \cite{VanRaamsdonk:2009ar,mvr2} about the emergence of spacetime in AdS/CFT. There, it was pointed out, based on the example of the eternal AdS black hole, that a classically connected spacetime can arise from quantum superpositions of spacetimes with two disconnected components. This phenomenon is apparent in the present setup: we have argued that typical microstates in the $H^d$ CFT thermal ensemble correspond to spacetimes similar to a Rindler wedge of AdS, but with the horizon replaced by something singular or non-geometrical. In the state (\ref{full}) describing pure AdS, we have a superposition of states $|E^L_i \rangle \otimes |E^R_i \rangle$, each of which can be interpreted as a disconnected pair of these microstate spacetimes. The quantum superposition (\ref{full}) represents pure global AdS spacetime, giving rise to the picture in Figure~\ref{superposition}. Compared to the earlier observations \cite{VanRaamsdonk:2009ar,mvr2}, a new feature is that (for this particular case) the microstate geometries contributing to the superposition are almost identical to pure AdS in their interior, but end rather abruptly at the place where the horizon would be in the connected version of the spacetime. Also, since there are many possible choices for complementary Rindler wedges in AdS, this example highlights the fact that there are many ways to decompose a given spacetime into a superposition of disconnected spacetimes.

Our conclusions about the geometry of Rindler microstates are similar to the ``fuzzball'' proposal of Mathur in that black hole microstates are geometries for which the horizon of the black hole has been replaced by something else. In our case, the ``something else'' may simply be some kind of lightlike singularity at which spacetime ends.\footnote{Note however, that some of the arguments we have used here are specific to the hyperbolic space CFT. In a similar discussion with $H^d$ replaced by $S^d$, we could argue in a similar way that the eternal black hole geometry behind the horizon has no relevance for black hole microstate spacetimes, but not that the microstates spacetimes are exactly the same as the black hole but with an abrupt end where the horizon would be.} Mathur has specifically proposed \cite{Mathur:2010kx, Mathur:2011wg} (based on a flat space limit of the observations in \cite{mvr2} about eternal AdS black holes) that a Rindler wedge of asymptotically flat space should have fuzzball microstates, and that empty space can be represented as a quantum superposition of disconnected geometries consisting of a pair of these Rindler fuzzballs. Our results confirm Mathur's suggestions in detail for the closely related case of empty AdS space. In this case, we have been able to give an explicit description of the Rindler-space theories (the hyperbolic space CFTs) and a (somewhat indirect) description of the fuzzball-geometries (the gravity duals of specific microstates of these hyperbolic CFTs).

\section{Rindler space results}
\label{rindler}

In this section, we consider a free massless scalar field theory on 1+1-dimensional Minkowski space and its alternative description based on the fields in a pair of complementary Rindler wedges. We prove that if a state factorizes into left and right components, i.e. if it is not entangled, its energy-momentum diverges on the Rindler horizon. This result suggests that ``AdS microstates'' discussed in the previous section are singular on the horizon of the AdS Rindler wedge. Interestingly, a divergent stress-energy on the boundary violates the AdS asymptotics, so the microstates cannot even be said to be asymptotically AdS! In the next section we complement this calculation with evidence that a state without entanglement represents a spacetime whose two parts have pinched off and disconnected from one another.

Consider a scalar field $\phi$ in two-dimensional Minkowski spacetime. We would like to divide this spacetime into a left and a right Rindler wedge. Defining $U = t-z$ and $V = t+z$, the right Rindler wedge is given by $U < 0 < V$. Because the dynamics of the left- and right-moving modes is independent and identical, we focus below on the right-moving sector, whose dynamics is independent of $V$.

A complete set of right-moving Rindler modes is
\begin{eqnarray}
\phi_{\lambda, R}(U) & = & \Theta(-U) \frac{1}{\sqrt{4\pi\lambda}}(-aU)^{i\lambda/a}~,
\label{A}\\
\phi_{\lambda, L}(U) & = & \Theta(U) \frac{1}{\sqrt{4\pi\lambda}}(aU)^{-i\lambda/a}~,
\label{B}
\end{eqnarray}
in terms of which, the field $\phi$ has an expansion:
\be
\phi(U) =
\int_0^\infty d\lambda ~\left ( b_{\lambda,R} \phi_{\lambda,R} +
b^\dagger_{\lambda,R} \phi_{\lambda,R}^*  \right ) ~+ ~
\int_0^\infty d\lambda ~\left ( b_{\lambda,L} \phi_{\lambda,L} +
b^\dagger_{\lambda,L} \phi_{\lambda,L}^*  \right )~. \label{D}\\
\ee
The relationship between the Minkowski vacuum and the Rindler
vacuum can be written as
\bear
&&|0,\mathrm{Mink}\rangle = {\cal U} |0,L\rangle|0,R\rangle~, \\ &&\mathrm{where}~~
{\cal U} = \prod_{\lambda>0}\exp\left( -\arctan(e^{-{\pi\lambda/a}})
(b_{\lambda,R}^\dagger b_{\lambda,L}^\dagger -
b_{\lambda,R} b_{\lambda,L})\right)~. \nn
\eear
For definiteness, we focus attention on the $UU$-component of the
stress-energy tensor.  Taking the expectation value of the stress-energy
tensor in the Minkowski vacuum as our reference point, we find that
the stress-energy in the Rindler vacuum is
\bear
\label{M}
T^{RL}_{UU} (U)&=& \langle 0,L|\langle 0,R|(\partial_U \phi(U))^2 |0,L\rangle|0,R\rangle ~-~
\langle 0,\mathrm{Mink}|(\partial_U \phi(U))^2 |0,\mathrm{Mink}\rangle \\
&=& -2~\int_0^\infty d\lambda ~\left [\beta_\lambda^2
\left ( |\partial_U \phi_{\lambda,R}|^2 +
|\partial_U \phi_{\lambda,L}|^2 \right )
~+~ 2 \alpha_\lambda\beta_\lambda \mathrm{Re}\left (
\partial_U \phi_{\lambda,L}\partial_U \phi_{\lambda,R} \right )
\right]~,
\nonumber
\eear
where $\alpha_\lambda = e^{\pi\lambda/a} \beta_\lambda = (1-e^{-2\pi\lambda/2})^{-1/2}$
are the Rindler coordinates Bogoliubov coefficients.

\begin{figure}
\centering
\includegraphics[scale=0.4]{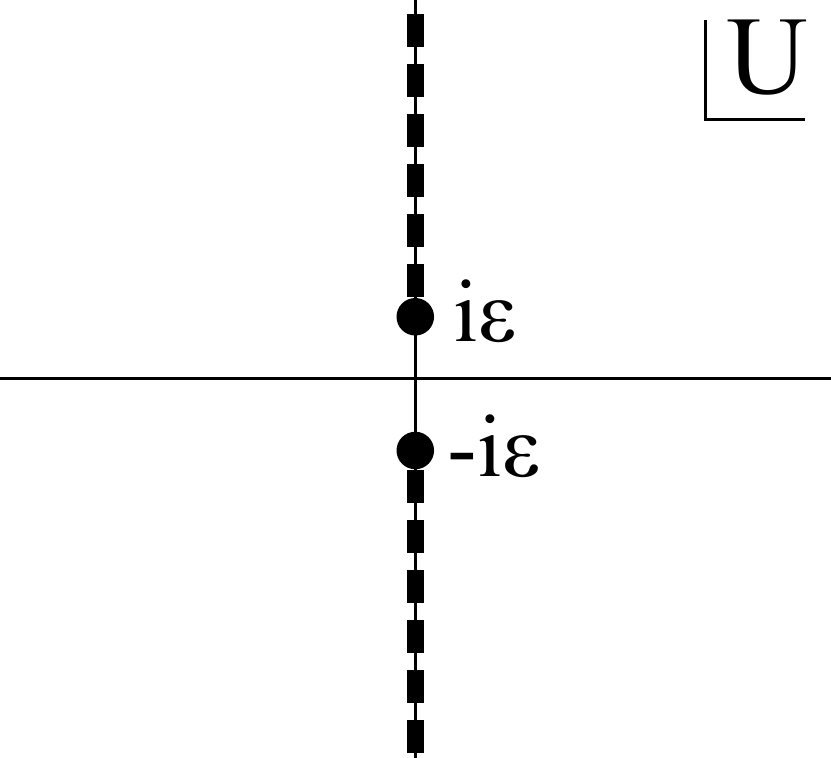}
\caption{ Position of cuts in the U-plane for formulas (\ref{I}) and
(\ref{J}).}
\label{uplane}
\end{figure}

At nonzero $U$, only the first term in equation (\ref{M}) is non-zero.
Using equations (\ref{A},\ref{B}), we get that
$T^{RL}_{UU} (U\neq0) = -1/(48\pi U^2)$.
To study the stress energy tensor at $U=0$, we follow the approach in
\cite{parentani}, and  regularize the operator
by replacing the modes $\phi_{\lambda,R/L}$ in equation (\ref{M})
with $\phi^\epsilon_{\lambda,R/L}$ given by
\bear
\phi^\epsilon_{\lambda,R}(U) &=& {1\over\sqrt{4\pi\lambda}}
{ \left(a(U-i\epsilon)\right)^{i\lambda/a}
- \left(a(U+i\epsilon)\right)^{i\lambda/a}
\over
e^{\pi\lambda/a} - e^{-\pi\lambda/a} }~, \label{I}
\\
\phi^\epsilon_{\lambda,L}(U) &=& {1\over\sqrt{4\pi\lambda}}
{e^{\pi\lambda/2} \left(a(U-i\epsilon)\right)^{-i\lambda/a}
-e^{-\pi\lambda/2} \left(a(U+i\epsilon)\right)^{-i\lambda/a}
\over
e^{\pi\lambda/a} - e^{-\pi\lambda/a} } ~.\label{J}
\eear
These expressions are valid and finite on the entire real line when
we place the cuts as shown Figure~\ref{uplane}.  Such placement of cuts
implies, for example, that for $U<0$, \be
(a(U-i\epsilon))^{i\lambda/a}  =
((e^{-i\pi})(-a(U-i\epsilon)))^{i\lambda/a}  =
e^{\pi\lambda/a}~(-a(U-i\epsilon))^{i\lambda/a} ~.
\ee
This ensures that $\phi^\epsilon_{\lambda,R/L}$ approach
 $\phi_{\lambda,R/L}$ for $\epsilon\rightarrow 0$.
We can now show that $\int_{-\infty}^\infty dU ~T^{RL}_{UU}$ is positive
and diverges for small $\epsilon$ like $\epsilon^{-1}$.  Thus, the Rindler
vacuum has a divergent stress-energy tensor on the boundary at $U=0$.

We now demonstrate that it is not possible to cancel this
singularity in the stress-energy tensor at the horizon in a general
separable state:\footnote{This state is more general than a tensor product of Fock space states of the left/right Rindler wedges, which takes the form
\be
\sum_{n=1}^\infty  \int \left (\prod_{i=1}^n d\lambda_i\right)
f_{n}(\lambda_1,\ldots,\lambda_n) \left ( \prod_{i=1}^n b^\dagger_{\lambda_i,L/R}\right ) |0\rangle_{L/R} \; .
\label{Y}
\ee
However, the typical states contributing to a finite temperature ensemble at infinite volume are of this more general form.}
\bear
|\Psi\rangle = \sum_{k=1}^\infty \sum_{n_i=1}^\infty \int \left (\prod_{i=1}^k d\lambda_i\right)
f_{(n_1,\ldots,n_k)}(\lambda_1,\ldots,\lambda_k) \left ( \prod_{i=1}^k (b^\dagger_{\lambda_i,L})^{n_i}\right ) |0\rangle_L  \\ \nonumber
\otimes \sum_{k=1}^\infty \sum_{n_i=1}^\infty \int \left (\prod_{i=1}^k d\lambda_i\right)
g_{(n_1,\ldots,n_k)}(\lambda_1,\ldots,\lambda_k) \left ( \prod_{i=1}^k (b^\dagger_{\lambda_i,R})^{n_i}\right )
 |0\rangle_R~.
\label{X}
\eear
The expectation value of the stress-energy tensor in this state is equal to that of the Rindler vacuum plus the additional contribution
\bear
T_{UU}^\Psi &=&\langle \Psi|(\partial_U \phi(U))^2 |\Psi\rangle ~-~ \nn
\langle 0,L|\langle 0,R|(\partial_U \phi(U))^2 |0,L\rangle|0,R\rangle  \\ &=&
\langle \Psi|:(\partial_U \phi(U))^2: |\Psi\rangle~,
\eear
where $::$ indicates normal ordering of Rindler raising and lowering operators.
Then
\begin{doublespace}
\bear
T_{UU}^\Psi= \int_0^\infty d\sigma_1 \int_0^\infty d\sigma_2 &\bigg [&
\langle \Psi| b_{\sigma_1,L} b_{\sigma_2,L} |\Psi\rangle  \:
\partial_U \phi_{\sigma_1,L} \partial_U \phi_{\sigma_2,L} ~+~ \mathrm{c.c.} \label{T1} \\ &&+~
2 \: \langle \Psi| b^\dagger_{\sigma_1,L} b_{\sigma_2,L} |\Psi\rangle \:
\partial_U \phi^*_{\sigma_1,L} \partial_U \phi_{\sigma_2,L} \label{T2}
\\ && +~
\langle \Psi| b_{\sigma_1,R} b_{\sigma_2,R} |\Psi\rangle \:
\partial_U \phi_{\sigma_1,R} \partial_U \phi_{\sigma_2,R} ~+~ \mathrm{c.c.} \label{T3} \\ &&+~
2 \:\langle \Psi| b^\dagger_{\sigma_1,R} b_{\sigma_2,R} |\Psi\rangle \:
\partial_U \phi^*_{\sigma_1,R} \partial_U \phi_{\sigma_2,R} \label{T4}\\ && +~
2 \: \langle \Psi| b_{\sigma_1,L} b_{\sigma_2,R} |\Psi\rangle \:
\partial_U \phi_{\sigma_1,L} \partial_U \phi_{\sigma_2,R} ~+~ \mathrm{c.c.}\label{T5} \\ &&+~
2 \: \langle \Psi| b^\dagger_{\sigma_1,L} b_{\sigma_2,R} |\Psi\rangle \:
\partial_U \phi^*_{\sigma_1,L} \partial_U \phi_{\sigma_2,R} ~+~ \mathrm{c.c.}\label{T6}~~\bigg ]~.
\eear
\end{doublespace} \noindent
Terms (\ref{T2}) and  (\ref{T4}) are non-negative everywhere, for example
\be
\int_0^\infty d\sigma_1 \int_0^\infty d\sigma_2
  \langle \Psi| b^\dagger_{\sigma_1,L} b_{\sigma_2,L} |\Psi\rangle \:
\partial_U \phi^*_{\sigma_1,L} \partial_U \phi_{\sigma_2,L}
~=~
\left | \int_0^\infty d\sigma~
 \partial_U \phi_{\sigma,L}~ b_{\sigma,L} |\Psi\rangle  \right |^2
\ee
and therefore their contribution to the stress-energy tensor cannot
cancel the Rindler vacuum singularity at $U=0$, which is also positive.

For terms (\ref{T1}), (\ref{T3}) and (\ref{T5}), consider the behavior
of the regularized stress-energy tensor near $U=0$,
$\int_{-\delta}^{\delta}dU T_{UU}^\Psi$ for small $\delta$.  For concreteness,
we focus on the (\ref{T1}), the argument for the other terms
being similar. Under the substitution
$U=\epsilon x$, we see that as $\epsilon \rightarrow 0$
\be
\int_{-\delta}^{\delta}dU~ \partial_U \phi_{\sigma_1,L} \partial_U \phi_{\sigma_2,L}~
\rightarrow ~\epsilon^{-1-i\sigma_1-i\sigma_2} \times ~(\mathrm{smooth~function~of~}\sigma_1\mathrm{~and~}
\sigma_2)~.
\ee
Due to the rapidly oscillating factor $\epsilon^{-i\sigma_1-i\sigma_2}$,
integrating over $\sigma_1$ and $\sigma_2$ in equation (\ref{T1}) will
give $\lim_{\epsilon \rightarrow 0} \epsilon \int_{-\delta}^{\delta}dU T_{UU}^\Psi = 0$.
The locus where the oscillations cancel, $\sigma_1+\sigma_2=0$, lies outside
the region of integration $\sigma_1>0, \sigma_2 >0$, so even if
$\langle \Psi| b_{\sigma_1,L} b_{\sigma_2,L} |\Psi\rangle$ were to contribute a
delta function $\delta(\sigma_1-\sigma_2)$ to the integrand, the integral would
remain zero.

The remaining term, (\ref{T6}), could give a non-zero contribution,
if $\langle \Psi| b^\dagger_{\sigma_1,R} b_{\sigma_2,L} |\Psi\rangle$ contributed
$\delta(\sigma_1-\sigma_2)$, as the rapidly
oscillating term takes the form $\epsilon^{\pm i(\sigma_1-\sigma_2)}$.
However, the separable form of our state $|\Psi\rangle$ does not allow for such
a delta-function term in $\langle \Psi| b^\dagger_{\sigma_1,R} b_{\sigma_2,L} |\Psi\rangle$.

Thus any separable state has a divergent stress-energy tensor on the boundary at $U=0$.
To cancel the singularity in the stress-energy tensor on the boundary,
we need an entangled state.  As an example, consider the Minkowski vacuum
written in the following suggestive and convenient form
\be
|\mathrm{Mink}\rangle = \prod_{\lambda>0}
\left [ {1\over \sqrt{Z_{\lambda}}}
\sum_{n=0}^\infty e^{-\pi n\lambda/a} {(b^\dagger_{\lambda,R})^{n}\over \sqrt{n!}}
{(b^\dagger_{\lambda,L})^{n}\over \sqrt{n!}} \right ]
|0,R\rangle |0,L\rangle~,
\label{N}
\ee
such that each wedge is in a thermal density matrix with the Rindler
temperature $T=a/2\pi$.
In this state, $\langle \mathrm{Mink}| b_{\sigma_1,R} b_{\sigma_2,L}| \mathrm{Mink}\rangle
=\delta(\sigma_1-\sigma_2) \:
e^{-\pi \sigma_1/a}/ (1-e^{-2 \pi \sigma_1/a}) = \delta(\sigma_1-\sigma_2) \:
\beta_{\sigma_1} \alpha_{\sigma_1}$.
Since $T_{UU}^{|\mathrm{Mink}\rangle} = -T^{RL}_{UU}$,
this delta-function contribution must cancel the divergence at $U=0$ precisely.

We have demonstrated than the singularity in the stress-energy tensor
at $U=0$ can only be cancelled in a state with entanglement between the right
and the left Rindler wedge.  Our discussion in Section \ref{microstates}
indicates that it should be possible to cancel the Rindler vacuum stress-energy
tensor in the interior of a Rindler wedge by adding Rindler quanta to
the Rindler vacuum.  To complete our discussion, we will now show that
we can achieve this to any desired accuracy with only a single quantum. Consider:
\be|\Psi_1\rangle = \int d\lambda f(\lambda)
b^\dagger_{\lambda,L} |0\rangle_L |0\rangle_R~,
\ee
where
\be
f(\lambda) = {e^{-(\lambda-\lambda_0)^2/(2\Delta^2)} \over \sqrt{2\pi}\Delta}~.
\ee
At nonzero $U$ we get (approximately, with $\Delta$ small enough)
\be
T_{UU}^{\Psi_1} = 2\left|\int d\lambda f(\lambda) \partial_U\phi_{L,\lambda}\right|^2 =
{\lambda_0 \over 2 \pi a^2 U^2 }
e^{-\Delta^2 (\ln(aU))^2/a^2}~.
\ee
For $(aU)$ in the interval $[e^{-a/\Delta},e^{a/\Delta}]$,
$T_{UU}^{\Psi_1}$ is approximately ${\lambda_0 \over 2 \pi a^2 U^2 }$.
By adjusting $\Delta$ and $\lambda$ appropriately, we can therefore
construct a state with arbitrarily small total stress-energy tensor $T_{UU}^{\Psi_1} + T^{RL}_{UU}$
inside the shaded region in Figure \ref{interior}(b).

\section{Effects of disentangling on geometry}
\label{disentanglement}

We have seen that entanglement is crucial for the description of pure AdS space in terms of a pair of hyperbolic space CFTs. To further highlight this, we consider in this section the effects on the bulk geometry of changing the amount of entanglement between the degrees of freedom in the two theories, which correspond to the two halves of the sphere in the original picture. This provides an explicit example of the ``disentangling experiment'' proposed in \cite{mvr2}.

As we have seen, the ``Rindler'' description of pure global AdS space is given by the state
\be
\label{fullstate3}
|0_M \rangle = {1 \over Z} \sum_i e^{- \frac{\beta E_i}{2}} |E^L_i \rangle \otimes |E^R_i \rangle \;
\ee
with temperature chosen as $\beta = 2 \pi R_H$. In this state, the degrees of freedom in the two hyperbolic space CFTs are entangled with each other. The claim in \cite{mvr2} was that if we change the state so that this entanglement is decreased, the dual spacetime should pinch off in the sense that the area of the bulk minimal surface separating the two halves should decrease and the distance between points in the two asymptotic regions should increase.

In the present context, we can decrease (or increase) the entanglement between the two sides by lowering (or raising) the temperature in the state (\ref{fullstate3}).\footnote{The entanglement can be quantified by the entanglement entropy $S = -\tr\, \rho_R \ln\rho_R$. This entropy is divergent, but can be regulated by introducing an ultraviolet cutoff in the theory. In this case, the difference between the entanglement entropy for two different states of the theory (considering the same pair of complementary spacetime regions) should be finite and regulator independent as the cutoff is removed.} In this case, each separate CFT on $H^d$ will be in a thermal state, corresponding in the bulk to an asymptotically AdS black hole (brane) with boundary geometry $H^d$. These black holes were described and interpreted in the AdS/CFT context in \cite{emparan}. The full state (\ref{fullstate3}) corresponds to the maximally extended version of these black holes with two asymptotic regions. By studying the geometry of a spatial slice of these black hole spacetimes as a function of $\beta$, we will see that the qualitative expectations in \cite{mvr2} are precisely realized in this explicit example.

\subsection{Review of the hyperbolic black holes}

In $d+2$ spacetime dimensions, the hyperbolic black hole geometry for temperature $T$ is described by the metric
\be
\label{metric}
ds^2 = -f(r) dt^2 + {dr^2 \over f(r)} + {r^2 \over l^2} (dH^d)^2\,
\ee
with
\bee
f(r) = {r^2 \over l^2} - {\mu \over r^{d-1}} -1 \; .
\eee
This has temperature
\bee
\beta = {4 \pi l^2 r_+ \over d r_+^2 - l^2 (d-1)}\,,
\eee
where $r_+$ is the horizon radius defined by $f(r_+) = 0$. The case $\mu = 0$ corresponds to the ``topological black hole'' that is a patch of pure AdS space. Both positive and negative values of $\mu$ are allowed, with the constraint that
\bee
\mu > \mu_{ext} = -{2 \over d-1}\left({d-1 \over d+1}\right)^{d+1 \over 2} l^{d-1} \; .
\eee
These coordinates cover the region exterior to the horizon, but the spacetime can be extended in the usual way to include a second asymptotic region (or more in the case $\mu < 0$). The causal structure is similar to the Schwarzschild-AdS black hole for $\mu>0$ and to the Reissner-Nordstrom AdS black hole for $\mu < 0$.

\subsection{Geometrical effects of changing the temperature/entanglement}

We would now like to compare the geometries for different values of $\mu$ (which controls the temperature/entanglement). Note that the boundary geometry is fixed; for all values of $\mu$ the metric takes the asymptotic form:
\be
\label{metric2}
ds^2 = -\frac{r^2}{l^2} dt^2 + {l^2  \over r^2}dr^2 + \frac{r^2}{l^2} (dH^d)^2
\ee
Thus, we can match the various spacetimes asymptotically by identifying points with the same $t$, $r$, and $H^d$ coordinates in the region of large $r$.

\subsubsection*{Distance across the spacetime}

\begin{figure}
\centering
\includegraphics[width=0.55\textwidth]{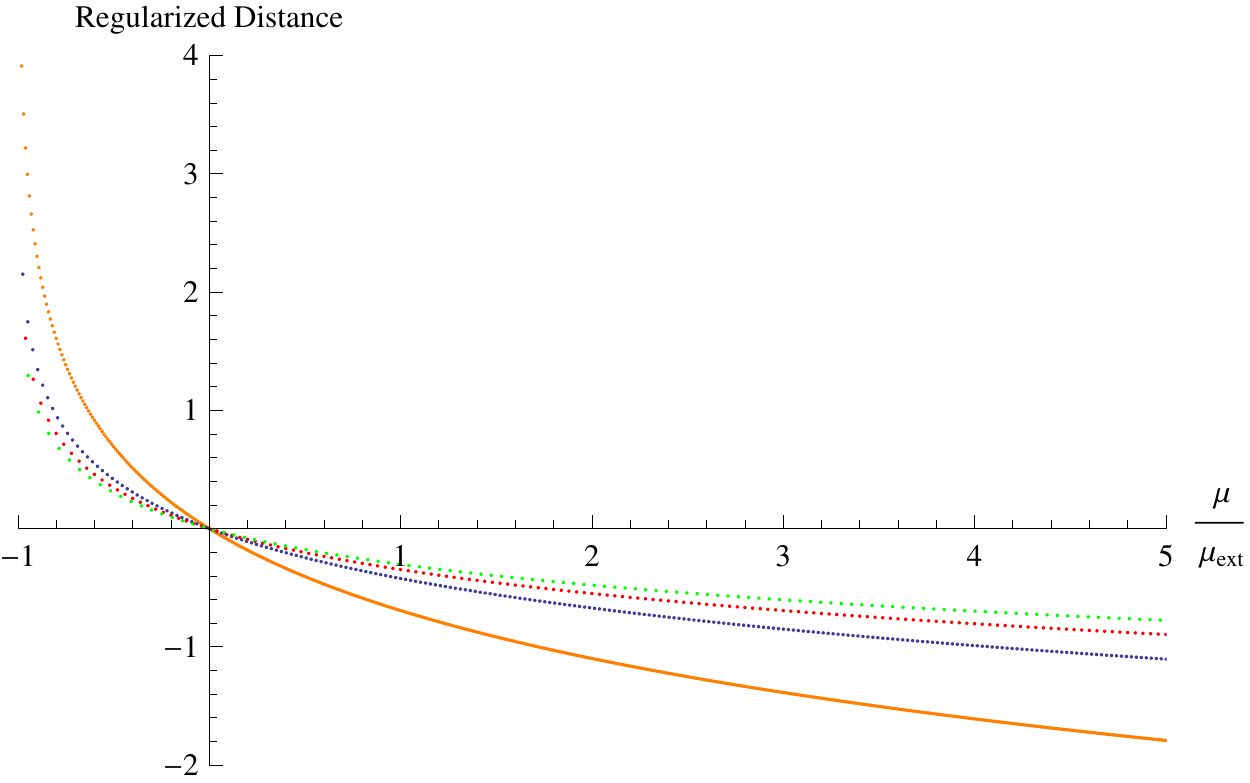}
\caption{Numerical integration of regularized distance for d=1, 2, 3, 4 (orange, blue, red, green, respectively).}\label{reg.dist.fig}
\end{figure}

First, we ask how the distance across the spacetime from one asymptotic region to the opposite one depends on $\mu$. Of course, the distance is infinite, but its deviation from the $\mu=0$ case of pure $AdS$ is finite and well defined. To compute this, we can choose some cutoff distance $R$. Then the distance across the spacetime on the $t=0$ slice (corresponding to the $\tau=0$ slice in global coordinates) at the origin of the hyperbolic space is
\bee
2 \int_{r_+}^R {dr \over \sqrt{f(r)}} \; .
\eee
Subtracting off the result for $\mu=0$ (with the same cutoff $R$) and taking the limit as $R \to \infty$ gives
\be\label{reg.dist}
L(\mu) - L(0) = 2 \int_{r_+(\mu)}^\infty dr \left\{ {1 \over \sqrt{{r^2 \over l^2} - {\mu \over r^{d-1}} -1}} - {1 \over \sqrt{{r^2 \over l^2}  -1}} \right\} - 2 \int_l^{r_+(\mu)} \left\{  {dr \over \sqrt{{r^2 \over l^2}  -1}} \right\}
\ee
This is finite, since the integrand in the first integral behaves as $1/r^{d+1}$ for large $r$. For $d=1$, we have explicitly that
\bee
\Delta L = - l \ln (1+\mu) .
\eee
Thus, the two sides of the spacetime get further apart as the entanglement between the corresponding degrees of freedom decreases. The same conclusion holds for other values of $d$ as indicated by a numerical evaluation of equation (\ref{reg.dist}) (see Figure~\ref{reg.dist.fig}). These results are consistent with the general expectations in \cite{mvr2}.

\subsubsection*{Areas of minimal surfaces}

We can similarly look at the areas (i.e. $d$-dimensional volumes) of minimal surfaces in the spacetime. We first consider the surface $r=r_+$ that divides the spacetime in half and forms the horizon of the hyperbolic black hole. The area of this is infinite, but we can look at the area per unit field theory volume as a function of $\mu$. This is proportional to $r_+^d$, where $r_+$ is related to $\mu$ by $\mu = r_+^{d-1}(r^2/l^2-1)$ (monotonic for $\mu > \mu_{ext}$).  Thus, the area of the surface separating the two halves of the space increases monotonically as we increase the entanglement (e.g. for $d=1$, we get Area~$\propto \sqrt{\mu+1}$).

We can also look at the areas of other $t=0$ spacelike surfaces that approach smaller spherical regions of the boundary. These extremize the action
\be
\label{length2}
A = \text{Vol}(S^{d-1})\int dr\, r^{d-1} \sinh^{d-1}\!u(r) \,\sqrt{{1 \over \frac{r^2}{l^2}
- \frac{\mu}{r^{d-1}}-1} + r^2 \left({du \over dr}\right)^2} \; .
\ee
For $d=1$, the action simplifies to
\be
\label{length}
L = \int dr\, \sqrt{{l^2 \over r^2 - l^2 (1+ \mu)} + r^2 \left({du \over dr}\right)^2} \; .
\ee
In this case, the path $u(r)$ must satisfy
\bee
{d \over dr} \left\{ {r^2 {du \over dr} \over \sqrt{{l^2 \over r^2 - l^2 (1+ \mu)}
+ r^2 \left({du \over dr}\right)^2}} \right\} = 0
\eee
Assuming $dr/du=0$ at some $r=r_{min}$, we have
\bee
{du \over dr} = {1 \over r} {l \over \sqrt{r^2 - l^2 (1 + \mu)}}{r_{min} \over \sqrt{r^2 - r_{min}^2}}
\eee
Setting the origin of hyperbolic space $u=0$ at this $r_{min}$, we find that the asymptotic value of $u$ (which we call $u_0$) as $r \to \infty$ is:
\bee
u_0 = {\ln {r_{min} + l \sqrt{1+ \mu} \over r_{min} - l \sqrt{1 + \mu}} \over 2 \sqrt{1+ \mu}}\;
\eee
Inverting the relationship, we obtain:
\bee
r_{min} = l \sqrt{1+\mu} \,\coth{(u_0 \sqrt{1 + \mu})}
\eee
From expression (\ref{length}), the length of such a curve in the region $r<R$ of spacetime is:
\begin{eqnarray}
l(\mu,R) &=& 2l \int_{r_{min}}^R {dr \over \sqrt{r^2 - l^2(1 + \mu)}}{r \over \sqrt{r^2- r_{min}^2}} \\
&=& 2l \ln \frac{\sqrt{R^2-r_{min}^2}+\sqrt{R^2-l^2(1+\mu)}}{\sqrt{r_{min}^2 - l^2(1+\mu)}} \nonumber
\end{eqnarray}
We can again subtract off the value for $\mu=0$ and take the limit as $R \to \infty$ to obtain the finite result:
\bee
l(\mu) - l(0) = 2l \ln \frac{\sinh( u_0\sqrt{1+\mu})}{\sinh(u_0) \sqrt{1+\mu}} \; .
\eee

\begin{figure}
\centering
\includegraphics[width=0.55\textwidth]{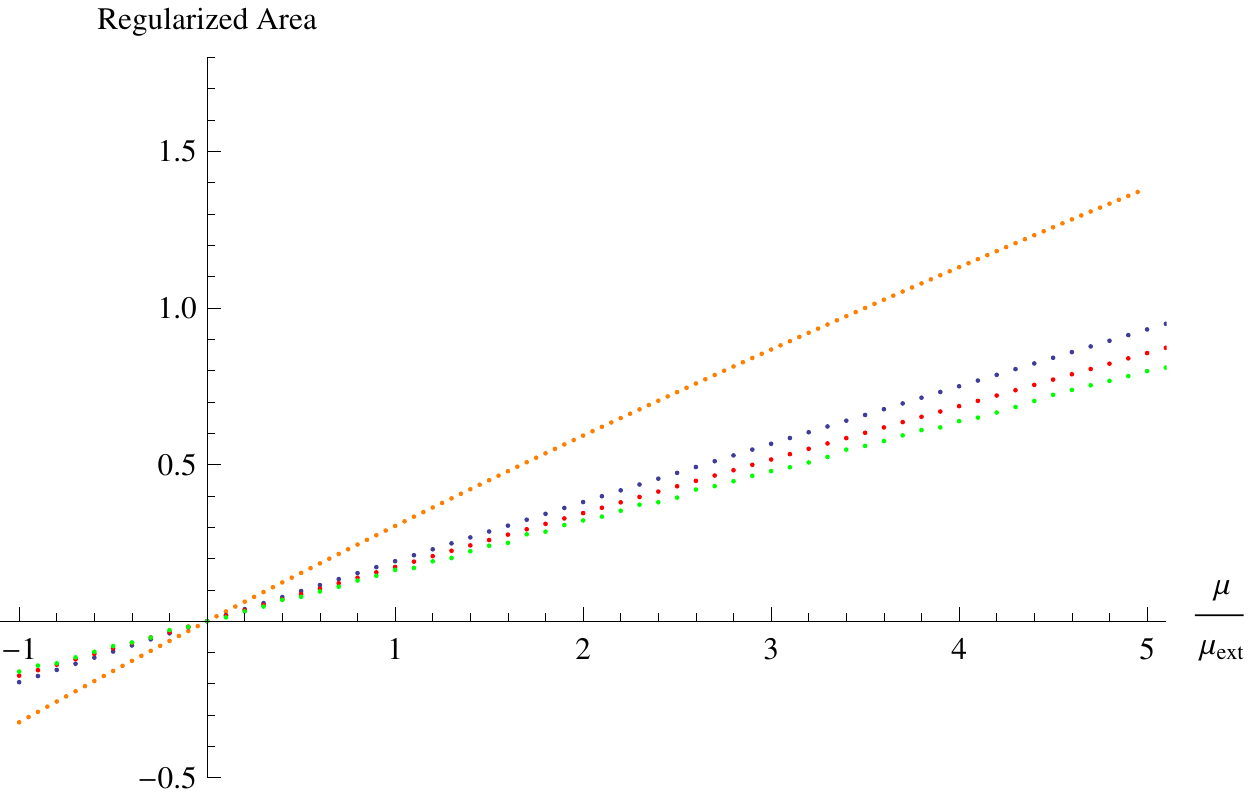}
\caption{Numerical integration of regularized distance with $u_0=1$ for d=1, 2, 3, 4 (orange, blue, red, green, respectively).}\label{reg.area.fig}
\end{figure}

We see that the area separating the two regions decreases as we lower the temperature (hence decreasing the entanglement entropy). The higher-dimensional versions can be tackled numerically, and we see that essentially the same pattern is repeated for all cases (Figure~\ref{reg.area.fig}). These results, together with the distance across spacetime, provides a realization of the ideas in \cite{mvr2}, that is, as entanglement entropy decreases, the two wedges of spacetime pinch off from each other.

\subsection{CFT on $S^d$ interpretation of the $H^d$ states at different temperatures}
\label{sdhd}

At temperature $T = (2 \pi R_H)^{-1}$, the state (\ref{fullstate3}) maps back to the vacuum state of the $S^d$ CFT, so the energy density is constant on the sphere (equal to the Casimir energy density). For other temperatures, the energy density is spatially constant and time-independent in the hyperbolic space picture, but not in the $S^d$ description. In this section, we determine explicitly the stress-energy tensor on $S^d$ for the state corresponding to (\ref{fullstate3}) at an arbitrary temperature.

For the states corresponding to hyperbolic black holes at various temperatures, the stress-energy tensor in the dual CFT on hyperbolic space with metric
\be
\label{hyp}
ds^2 = - dT^2 + R^2 (du^2 + \sinh^2\!u\, d \Omega^2_{d-1})
\ee
is given by \cite{emparan}
\be
\label{hypstress}
\langle T^\mu {}_\nu \rangle = {1 \over 16 \pi G l} \left( \epsilon_d + {\mu \over l^{d-1}}\right)
\,\diag(-d,1,\dots,1),
\ee
where
\bee
\epsilon_d = {2(d!!)^2 \over (d+1)! d}
\eee
for odd $d$ and zero for even $d$. For these states, we can map back to states of the CFT on $S^d$. In this case, we have the metric on the region $D$ is conformally related to the hyperbolic space metric:
\bee
g_{\mu \nu}^{sphere} = e^{2 \phi} g_{\mu \nu}^{hyp}
\eee
Hence, we conclude that \cite{Birrell:1982ix}
\bee
(\langle T_{sphere} {}^\alpha {}_\beta \rangle_{\mu} -  \langle T_{sphere} {}^\alpha {}_\beta \rangle_{\mu=0}) = e^{- (d+1) \phi}(\langle T_{hyp} {}^\alpha {}_\beta \rangle_{\mu} - \langle T_{hyp} {}^\alpha {}_\beta \rangle_{\mu=0})\,,\label{relatet}
\eee
where the $\mu=0$ state corresponds to the vacuum of the field theory on the sphere.

Starting from the metric (\ref{hyp}) for hyperbolic space times time, the change of coordinates
\begin{eqnarray}
\tan(\tau/R) &=& {\sinh(T/R) \over \cosh u} \\
\tan\theta &=& {\sinh u \over \cosh(T/R)}
\end{eqnarray}
gives
\bee
ds^2=e^{2 \phi}\big( -d\tau^2 + R^2(d \theta^2 + \sin^2\theta\, d \Omega^2_{d-1})\big)
\eee
with
\bee
e^{2 \phi} = {\cos^2(\tau/R) + \sin^2\theta \over \cos^2(\tau/R) - \sin^2\theta} \; .
\eee
In these coordinates, the stress tensor is
\bee
\langle T^\mu {}_\nu \rangle = {1 \over 16 \pi G l} \left( \epsilon_d + {\mu \over l^{d-1}}\right) \big( \diag(1,1,\dots,1) - (d+1) \;  {\bf M} \big)\,,
\eee
where
\bee
{\bf M} =  \left( \ba{ccc}
{1 \over 1 - \tan^2\!\theta\, \tan^2(\tau/R)}
& {\sin\!\theta\, \sin(\tau/R) \cos\!\theta\, \cos(\tau/R) \over 1 - \sin^2\!\theta - \sin^2(\tau/R)}
& \cr - {\sin\!\theta\, \sin(\tau/R) \cos\!\theta\,\cos(\tau/R) \over 1 - \sin^2\!\theta\, - \sin^2(\tau/R)}
& {1 \over 1 - \cot^2\!\theta\, \cot^2(\tau/R)} &  \cr
& & {\bf 0} \ea \right)
\eee
Using (\ref{relatet}), we find that the stress tensor for the corresponding state of the CFT on the domain of dependence of the half-sphere (i.e. the region $D$) with metric
\bee
ds^2 =  -d\tau^2 + R^2\big(d \theta^2 + \sin^2\theta\, d \Omega^2_{d-1}\big)
\eee
is
\be
\label{Tres}
\langle T_{sphere} {}^\alpha {}_\beta (\mu)\rangle -  \langle T_{sphere} {}^\alpha {}_\beta \rangle_{vac} = {\mu \over 16 \pi G l^d} \left( {\cos^2(\tau/R) + \sin^2\theta \over \cos^2(\tau/R) - \sin^2\theta} \right)^{(d+1)/2} \big({\bf 1} - (d+1) {\bf M} \big)\,.
\ee
As an example, the energy density (minus the Casimir energy) is given by
\bee
T^{00} - T_{vac}^{00} = {\mu \over 16 \pi G l^d} \left( {\cos^2(\tau/R) + \sin^2\theta \over \cos^2(\tau/R) - \sin^2\theta} \right)^{(d+1)/2} \left( {d +1\over 1 - \tan^2\!\theta\, \tan^2(\tau/R)} -1\right)
\eee
This diverges on the lightlike boundary of the causal development region of the half-sphere, $\tau/R = \pm (\pi/2 -\theta)$. At $\tau=0$ the energy density
\bee
T^{00} - T_{vac}^{00} = {d\mu \over 16 \pi G l^d} \left( {1 + \sin^2\theta \over 1 - \sin^2\theta} \right)^{(d+1)/2}
\eee
diverges at the equator $\theta = \pi/2$ and this singularity propagates forward and backward in time along the light sheets. We note also that for $\mu<0$ the $\tau=0$ energy density is negative away from the equator.\footnote{The negative energy density here is a well-known possibility, which illustrates how the weak and null energy conditions may be violated. The negative energy should be thought of as a Casimir-type vacuum energy. Certain inequalities restrict the extent of such negative energy densities. They can be used to prove averaged versions of the energy conditions in certain situations (see e.g. \cite{Ford:1994bj}).} However, since the total energy on the sphere must be higher than for the vacuum state of the CFT, there must be a singular positive contribution to the energy density at the equator such that the total energy on the sphere (relative to the vacuum energy) is positive.

While we have focused in this section on some CFT with a gravity dual, the stress-energy tensor for an arbitrary conformal field theory at finite temperature on $H^d \times R$ is determined by homogeneity and conformal invariance (tracelessness) to be
\be
\label{genstress}
\langle T^\mu {}_\nu \rangle = f(2 \pi R_H T) \,\diag(-d,1,\dots,1) ,
\ee
proportional to (\ref{hypstress}) that was our starting point. Thus, the result (\ref{Tres}) holds in general, with the replacement
\bee
{\mu \over 16 \pi G l^d} \to f(2 \pi R_H T) - f(1) \; .
\eee
In particular, except for the special temperature $T = (2 \pi R_H)^{-1}$ that corresponds to the vacuum state on $S^d$, the stress energy tensor is singular at the boundary of the domain of dependence of the half-sphere.

\section{Comments on generalization to cosmological spacetimes}
\label{cosmology}

In the introduction, we recalled various qualitative similarities between Rindler patches of AdS and patches accessible to observers in cosmological spacetimes. Based on these similarities, it seems plausible that the description of physics inside a cosmological horizon should be in terms of a density matrix for some degrees of freedom. However, both the details of the patch geometry and the local spacetime dynamics are different in cosmological examples. In this section, we offer a few comments on how the holographic description might be modified in going from the case of accelerated observers in AdS to the case of observers in cosmological spacetimes\footnote{For some other approaches to this question, see \cite{strominger,gm,Alishahiha:2004md,fssy,bf}.}.

A characteristic feature of asymptotically AdS spacetimes not present in the cosmological examples is the AdS boundary. The patches accessible to an observer in de Sitter space or other homogeneous spacetimes with accelerated expansion are bounded by the cosmological horizon and have finite spatial volume. In the AdS case, all the patches we have described have infinite spatial volume since they include a boundary region. We know that the boundary region is tied to the UV degrees of freedom in the field theory. Thus, we might guess that patches of AdS without the boundary region are described by a reduced density matrix for a subset of field theory degrees of freedom that excludes the UV degrees of freedom; such density matrices have been considered recently in \cite{Balasubramanian:2011wt}. For a CFT on $S^d$, excluding the UV degrees of freedom (e.g. spherical harmonic modes of the fields with angular momenta above a certain cutoff) leaves us with a finite number of degrees of freedom, those of a large $N$ matrix model with a finite number of matrices. Thus, a description for finite volume patches might be via mixed states for a large $N$ matrix model, where these matrix model degrees of freedom are entangled with (and perhaps interacting with) other degrees of freedom associated with the rest of the spacetime.\footnote{We are not suggesting that arbitrarily small or localized patches of spacetime can be associated with some particular degrees of freedom, only that certain patches may be associated with certain mixed states of a model with a finite number of degrees of freedom.} A very similar conclusion was reached by Susskind in \cite{Susskind:2011ap} for independent reasons.

At a more detailed level, in order to describe local bulk physics characteristic of a spacetime with positive, rather than negative cosmological constant, we should expect that the Hamiltonian associated with time evolution in some patch should be different from one describing patches of AdS.  Short of providing a specific suggestion here, we only observe that for a particular geodesic trajectory in AdS, flat, and de Sitter space, other geodesic trajectories respectively accelerate towards, move away at constant velocity, or accelerate away from this trajectory. In the context of matrix models, these behaviors can be put in ``by hand'' at the classical level by choosing positive, zero, or negative mass-squared terms for bosonic degrees of freedom. Thus, a completely speculative suggestion would be that the type of matrix model whose mixed states would describe physics in a patch of a spacetime with accelerated expansion might involve negative mass squared terms for the bosonic matrices.\footnote{A slightly more concrete motivation of this suggestion is as follows. Starting from the ${\cal N}=4$ SYM theory on $S^3$, a particular way to truncate to the IR degrees of freedom is to keep only the lowest spherical harmonic modes. This can be done in a way that preserves maximal supersymmetry, and the result is the Plane Wave Matrix Model, which has positive mass for all bosonic degrees of freedom. The density matrix for this model that arises starting from the vacuum of ${\cal N}=4$ SYM and tracing out the rest of the degrees of freedom should describe a patch of pure AdS. For flat spacetime, the most concrete proposals for dual descriptions involve limits of models for which the bosonic potential has many flat directions (e.g. the BFSS matrix model). It is from these flat directions (preserved at the quantum level) that the asymptotic flatness of the dual spacetime is supposed to emerge. Thus, our naive suggestion is realized in specific models for the AdS and flat cases.} We caution, however, that quantum effects typically dominate the effective potential in a matrix model; only for very special theories, typically with significant cancellations in the effective potential due to supersymmetry, do we expect any kind of dual spacetime picture to emerge. For an alternate (and more in-depth) discussion on how to modify CFT physics in order to describe de Sitter or FRW (rather than AdS) dynamics, see \cite{Dong:2010pm,Dong:2011uf}.

\section*{Acknowledgments}

We would like to thank Ted Jacobson, Don Marolf, Samir Mathur, Bill Unruh, Aron Wall and the participants of the KITP ``Bits, Branes, and Black Holes'' workshop for helpful discussions and comments. BC and MVR would like to thank the KITP for hospitality during the workshop. This work is supported in part by the Natural Sciences and Engineering Research Council of Canada, by the Canada Research Chairs Programme,  by the National Science Foundation under Grant No. NSF PHY11-25915, and by a grant from the Foundational Questions Institute (FQXi) Fund, a donor advised fund of the Silicon Valley Community foundation on the basis of proposal FQXi-RFP3-1010.

\appendix

\section{Coordinate Transformations}\label{Coordinate.tranformation}

In this appendix, we show that a conformal transformation between the boundary of some Poincare patch and Minkowski space maps diamond-shaped regions as in Figure~\ref{Dtorind1} to complementary Rindler wedges of Minkowski space. We also argue that there is another conformal transformation that maps one of these diamond-shaped regions to hyperbolic space times time.

Starting with the cylinder $S^d \times R$ in coordinates
\be\label{globalmetric}
ds^2=-dT^2+dR^2+\sin^2\!R\,d\Omega_{d-1}^2,
\ee
the change of coordinates
\be
\tan {T\pm R \over 2} = t \pm r
\ee
followed by the conformal transformation
\be
ds^2 \to \frac{1}{4} \left((r-t)^2+1\right)\left((r+t)^2+1\right) ds^2
\ee
takes the region $D_P = \{-\pi < T < \pi, R < \pi - |T|\}$ to Minkowski space with metric
\be
ds^2=-dt^2+dr^2+r^2d\Omega_{d-1}^2 \;.
\ee
The region $D_P$ forms the boundary of a Poincare patch in AdS. If $p_i$ and $p_f$ are any points on the past and future boundaries of $D_P$ (the past and future tips of a diamond-shaped region as in Figure~\ref{Dtorind1}), then the forward and backward lightcones from $p_i$ and $p_f$ divide $D_P$ into four regions, as in Figure~\ref{Dtorind1}. After the transformation to Minkowski space, the space is still divided into four regions by lightcones, but since $p_i$ and $p_f$ map to the infinite past and infinite future, these lightcones become intersecting lightlike planar hypersurfaces. After a Poincare transformation, these can be mapped to the surfaces $x=t$ and $x=-t$ that bound Rindler wedges of Minkowski space, with the region $D(p_i,p_f)$ (the region bounded by the forward lightcone from $p_i$ and the backward lightcone from $p_f$) mapping to one of the wedges.
As an example, the transformation above, without any additional Poincare transformation, maps the domain of dependence of the $\theta < \pi/2$ hemisphere of the $T=0$ sphere to the Rindler wedge $x>0, |t| < x$.

To map $D(p_i,p_f)$ to $H^d \times R$ using a conformal transformation, we can combine a map $D(p_i,p_f)$ to a Rindler wedge of Minkowski space as above, with a map back to a region $|T|<T_0, R < T_0 - |T|$ (the causal development of a ball in the $T=0$ sphere), with a third conformal transformation (given explicitly in \cite{Casini:2011kv}) to $H^d \times R$. We note in particular \cite{Casini:2011kv} that in the map from the Rindler wedge to $H^d \times R$, the Rindler Hamiltonian maps to the generator of time translations.

\end{document}